\documentclass[10pt,letterpaper]{article}
\usepackage[top=0.85in,left=2.75in,footskip=0.75in,marginparwidth=2in]{geometry}

\usepackage[utf8]{inputenc}

\usepackage{cite}

\usepackage{nameref,hyperref}

\usepackage[right]{lineno}

\usepackage{microtype}
\DisableLigatures[f]{encoding = *, family = * }

\usepackage{hyperref}

\usepackage[T1]{fontenc}
\usepackage{amsmath}
\usepackage[sort&compress,sectionbib,numbers]{natbib}
\raggedright
\usepackage{changepage}

\usepackage[aboveskip=1pt,labelfont=bf,labelsep=period,singlelinecheck=off]{caption}

\makeatletter
\renewcommand{\@biblabel}[1]{\quad#1.}
\makeatother

\usepackage{lastpage,fancyhdr,graphicx}
\usepackage{epstopdf}
\fancyheadoffset[L]{2.25in}
\fancyfootoffset[L]{2.25in}

\usepackage{color}

\definecolor{Gray}{gray}{.25}

\usepackage{graphicx}

\usepackage{sidecap}

\usepackage{wrapfig}
\usepackage[pscoord]{eso-pic}
\usepackage[fulladjust]{marginnote}
\reversemarginpar

\begin{document}
\vspace*{0.35in}

\begin{flushleft}
{\Large
\textbf\newline{Assessing the forensic value of DNA evidence from Y chromosomes and mitogenomes}
}
\newline
\\
Mikkel M Andersen\textsuperscript{1,2,*},
David J Balding\textsuperscript{3,4}
\\
\bigskip
\bf{1} Department of Mathematical Sciences, Aalborg University, Aalborg, Denmark; mikl@math.aau.dk
\\
\bf{2} Section of Forensic Genetics, Department of Forensic Medicine, University of Copenhagen, Copenhagen, Denmark
\\
\bf{3} Melbourne Integrative Genomics, University of Melbourne, Melbourne, Australia; dbalding@unimelb.edu.au
\\
\bf{4} Genetics Institute, University College London, London, United Kingdom
\\
\bigskip
* Correspondence: mikl@math.aau.dk

\end{flushleft}

\section*{Abstract}
 Y-chromosomal and mitochondrial DNA profiles have been used as evidence in courts for decades, yet the problem of evaluating the weight of evidence has not been adequately resolved. Both are lineage markers (inherited from just one parent), which presents different interpretation challenges compared with standard autosomal DNA profiles (inherited from both parents), for which recombination increases profile diversity and weakens the effects of relatedness. We review approaches to the evaluation of lineage marker profiles for forensic identification, focussing on the key roles of profile mutation rate and relatedness. Higher mutation rates imply fewer individuals matching the profile of an alleged contributor, but they will be more closely related. This makes it challenging to evaluate the possibility that one of these matching individuals could be the true source, because relatedness may make them more plausible alternative contributors than less-related individuals, and they may not be well mixed in the population. These issues reduce the usefulness of profile databases drawn from a broad population: the larger the population, the lower the profile relative frequency because of lower relatedness with the alleged contributor. Many evaluation methods do not adequately take account of relatedness, but its effects have become more pronounced with the latest generation of high-mutation-rate Y profiles.

\section{Weight of evidence for lineage marker profiles}
\label{sec:intro}

Standard DNA profiles use autosomal DNA, inherited from both parents. We focus here on DNA profiles obtained from the Y chromosome and the mitochondrial genome (mitogenome/mtDNA), which are inherited only from the father and from the mother, respectively. Because of this uniparental inheritance over generations, these DNA profiles are called lineage markers. We outline the forensic value of lineage markers in general, give a brief, historical review and critique of evaluation methods and make recommendations for improved practice. Our key message is that the high mutation rates of the latest generation of Y-chromosome short tandem repeat (STR) profiles have effects that exaggerate the deficiencies of previous methods of analysis, but understanding these effects highlights ways forward.

We focus on the simplest scenario in which there is a good-quality, single-contributor DNA profile (either Y or mitogenome) obtained from an evidence sample, and a matching reference profile from a known individual $Q$ who is alleged to be source of the evidence sample (sometimes $Q$ is called the person of interest, PoI). Hypotheses of interest are:
\begin{align*}
H_Q&: \mbox{the evidence profile came from } Q\\
H_X&: \mbox{the evidence profile came from } X, 
\end{align*}
where $X$ is an alternative to $Q$ as source of the DNA whose profile is not available. The likelihood ratio (LR) comparing the strength of the DNA profile evidence for $H_Q$ relative to $H_X$ is then
\begin{equation} 
 \mbox{LR}(X,Q) = \frac{\mbox{P(profile evidence} \mid H_Q)}{\mbox{P(profile evidence} \mid H_X)} = \frac{1}{\mbox{P}(X \mbox{ has profile } q)}, \label{eq:LR}
\end{equation}
where $q$ is the profile of $Q$. We omit background information and other evidence from the notation, see \citep{WoE2nd} for a discussion. The denominator of \eqref{eq:LR} is a probability for the unknown profile of $X$, given the observed $q$ and possibly a database of profiles.

\subsection{The effect of relatedness and mutation rate on the LR}

For lineage markers, the relatedness of two individuals is fully captured by a single number, $G$, of generations (or germline transfers, or meioses) that separate $X$ and $Q$, following either female-only or male-only ancestors. Degree-1 ($G=1$) relative pairs are parent/offspring, $G=2$ for siblings and grandparent/grandchild, while $G=3$ for avuncular relationships such as aunt/niece as well as great-grandparent/great-grandchild. As $G$ increases, the relatedness becomes less likely to be known, but there is a female-lineage $G$ and a male-lineage $G$ for all pairs of individuals, ``unrelated'' only means that $G$ is large and/or unknown.

Given $G$ for $X$ and $Q$, we can approximate \eqref{eq:LR} by
\begin{equation} 
 \mbox{LR}(X,Q) = \frac{1}{(1-\mu)^G}\label{eq:LRg}
\end{equation} 
where $\mu$ is the profile mutation rate\footnote{For $L$ loci each with mutation rate $\mu_l$, mutation events are usually considered to be independent across loci so that $\mu = 1 - \prod_{l=1}^L (1-\mu_l)$, which is less than (but usually close to) the sum of the $\mu_l$.}. This formula is not exact for at least two reasons. Firstly, it assumes that the profile mutation rate is constant over generations and independent of the current profile state. In fact, accurate estimates are hard to obtain but the STR mutation rate is likely to depend on the allele sequence (including its length and presence of a partial repeat) \cite{Jochens2011}. However, these effects are relatively unimportant for the total mutation rate over many loci. Secondly, it is based on assuming no mutations in the lineage path connecting $X$ and $Q$, whereas a match can also arise following an even number of mutations between $X$ with $Q$ such that the effect of each mutation is reversed by another mutation. However, profiles consisting of many loci and with multiple possible mutation events at each locus are very unlikely to match if there is any mutation between $X$ and $Q$ \citep{Brenner14, AndersenPLOSGEN2017}.

For $G$ unknown, in place of \eqref{eq:LRg} we have (see also \citep{Caliebe2018}):
\begin{equation} 
 \mbox{LR}(X,Q)=\frac{1}{\sum_{g=1}^\infty(1-\mu)^g\mbox{P}(G=g)}\qquad\mbox{where}\quad\sum_{g=1}^\infty\mbox{P}(G=g)=1.\label{eq:mp-meioses}
\end{equation}
Equation \eqref{eq:mp-meioses} requires a probability distribution for $G$, which can be informed by population genetic models and by the available information about alternative possible sources of the DNA profile. We introduced $X$ as a specific alternative individual, but in practice there are usually many alternative sources of the DNA. In either case, P$(G=g)$ can be interpreted as the probability that the unknown alternative source of the DNA is a degree-$g$ relative of $Q$. If, for example, it is known that all of the degree-$g$ relatives of $Q$ are excluded as possible sources, then $P(G=g) = 0$ in \eqref{eq:mp-meioses}\footnote{Note that after setting $P(G=g) = 0$ it must still be true that $\sum_{g=1}^\infty\mbox{P}(G=g)=1$, meaning that the distribution potentially has to be renormalised.}. 

\begin{table}[htb]
\centering
\begin{tabular}{|l l r l|}
 \hline
&& \textbf{Profile mutation rate} & \\ 
\textbf{Type} & \textbf{Markers} & [\% per generation] &\textbf{Reference}\\ \hline
 Ychr & PowerPlex Y (12 loci) & 2.5\% & \cite{AndersenPLOSGEN2017, malan} \\
 Ychr & Yfiler (17 loci) & 4.4\% & \cite{AndersenPLOSGEN2017, malan} \\ 
 Ychr & PowerPlex Y23 (23 loci) & 8.3\% & \cite{AndersenPLOSGEN2017, malan} \\
 Ychr & Yfiler Plus (27 loci) & 13.5\% & \cite{AndersenPLOSGEN2017, malan} \\
 Mitogenome & 16,070 sites & 0.3\%-1.9\% & \cite{Rieux2014, AndersenPLOSGEN2018} \\
 Mitogenome & 16,494 sites & 0.4\%-2.3\% & \cite{Oversti2017, AndersenPLOSGEN2018} \\ 
 \hline
\end{tabular}
\caption{Estimates of lineage marker profile mutation rate (= the probability that parent and child have non-matching profiles). These estimates are obtained assuming that the mutation probability does not depend on the current profile state. The per-year mitogenome mutation rates in \citep{Rieux2014, Oversti2017} have been multiplied by a generation time of 25 years.}
\label{tab:mut}
\end{table}

Although we can rarely compute it accurately, \eqref{eq:mp-meioses} tells us how to evaluate a lineage marker profile match: we need to assess, given the known circumstances, the probability that an alternative source of the DNA has relatedness $g$ with the alleged source $Q$, and weight this probability by $(1-\mu)^g\approx e^{-g\mu}$, which means that individuals with $g\gg1/\mu$ contribute little to the LR.

Most presentations of lineage marker DNA profile evidence mention something like ``maternally-related individuals will share the same mitochondrial DNA profile'' and then proceed to assess weight of evidence assuming that $X$ and $Q$ are unrelated. Such a statement conveys little and is potentially misleading, because we are all maternally related and we are also all paternally related, what is important is the degree $g$ of the relatedness (or that $g\gg1/\mu$). 

\subsection{The role of databases in evidence evaluation}

Most methods reviewed below make some use of a database of profiles. When multiple databases are available, the one with ancestry closest to that of $Q$ should usually be chosen \citep{SteeleSJ2014,RecomUK}, unless the database size is very small and there is an alternative database of larger size drawn from a population with similar ancestry.

Databases are usually not random samples \citep{AndersenPLOSGEN2017, ISFG2020}, and not drawn from the population relevant to a specific case. For lineage markers, the important role of relatedness implies that the way the database is sampled can have a big impact on inferences. On the one hand, databases may be sampled in a way that includes more sets of related individuals than would the case for random sampling, for example because the close relatives of a suspected contributor may also be suspects. On the other hand, some databases may implement a policy of excluding close relatives.

For Y and mitogenome the most important international databases, both highly respected for their data quality \cite{EmpopQC}, are available online at \url{www.YHRD.org} \citep{YHRD15} and \url{www.EMPOP.online} \citep{empop, ISFGmtDNA}. In July 2021 EMPOP had 48,572 mtDNA sequences (of which 4,289 cover the entire mitogenome, the rest span some or all of the control region, which is the most variable part of the mitogenome \cite{ISFGmtDNA}). YHRD had 337,449 minimal (8 STR loci) profiles, of which 97,087 were 27-locus Yfiler Plus profiles. Both databases contain samples from multiple worldwide populations, in some cases including several subpopulations.

Equation \eqref{eq:mp-meioses} has profound implications for the role of databases in computing the LR. For any of the lineage markers in Table~\ref{tab:mut}, the space of possible profiles is so vast that it is extremely unlikely that two distinct lineages will generate the same (complete) profile by neutral mutations. Except for partial profiles with few loci observed (Section \ref{sec:partial-profiles}), the possibility of a ``chance'' match between distantly related individuals is negligible relative to the much more likely event of a match between pairs of individuals who are related closely enough to generate the match, but distantly enough that the relatedness is not recognised. Relatedness also affects match probabilities for autosomal profiles, but because of recombination its effect is much less important except for monozygotic twins \citep{TvedebrinkTwins2015}. 

For the mitogenome, $\mu$ is estimated to be around 1 per 70 generations (Table~\ref{tab:mut}). Thus, alternative sources $X$ will be almost certainly be female-line relatives of $Q$ with $G$ up to a few hundred \citep{AndersenPLOSGEN2018}. This is greatly beyond the known relatives of $Q$, but is still closer than a random pair of individuals in a large population, who are typically separated by thousands of generations \citep{AndersenPLOSGEN2018}. Mutation rates for older Y-profile kits are a small multiple of the mitogenome rate, but more recent Y-profile kits can have mutation rate an order of magnitude higher: around 1 per 7.5 generations for Yfiler Plus profiles (Table~\ref{tab:mut}). In that case profile matches occur between pairs of individuals separated by at most a few tens of generations \citep{AndersenPLOSGEN2017}.

Therefore the majority of profiles in broadly-defined databases such as YHRD or EMPOP are drawn from individuals that are too distantly related to $Q$ to be relevant to a particular case. The frequency of $q$ can depend sensitively on the choice of population, according to its average relatedness with $Q$. Alternative sources of the DNA may be concentrated in the same subpopulation as $Q$, defined by geographical origin and/or social factors such as ethnicity and religion. A relevant consideration is that if $Q$ is in fact not the source of the DNA, the false allegation may in part be due to $Q$ resembling the true source in some characteristics such as appearance, place of residence, or social background. Similarity on these characteristics tends to increase P$(G=g)$ for small values of $g$. Therefore, a broadly-defined database population can lead to artificially low match probability estimates.

\section{Review of evaluation methods}
\label{sec:methods}

Most methods have not addressed the fundamental effects of relatedness, mutation rate and database sampling frame discussed above. We will review methods assuming, as most authors have done, that the available database is appropriate, and return to these issues in the Discussion.

\subsection{Adjusted database counts}
\label{sec:adjust}

The denominator of the LR \eqref{eq:LR}, is often assumed to equal the \textsl{match probability}, $\pi_q$, the relative frequency of $q$ in a population of alternative sources of the DNA. Many approaches aim to estimate $\pi_q$ based on the count $k_q$ of $q$ in a database of size $n$. The database relative frequency $k_q/n$ has long been recognised as an unsatisfactory estimator of $\pi_q$, because we often have $k_q=0$ and yet $Q$ has the profile and so $\pi_q>0$.

Further, the requirement to avoid overstatement of evidence converts here to preferring a "conservative" over-estimate of $\pi_q$ rather than an under-estimate. Excessive conservativeness wastes information, and our goal should be to evaluate the evidence as accurately as possible, while taking care to guard against being anti-conservative. However, the latter requirement is difficult to satisfy, and there is no agreement about to what extent we should seek to eliminate any risk of being anti-conservative.

Various adjustments to $k_q$ and $n$ have been proposed to avoid zero estimates and introduce an upward bias. We will discuss them in order of increasing conservativeness as measured by $\hat{\pi}_q$ when $k_q=0$.

\subsubsection{Adjustment based on the database frequency spectrum}
\label{sec:spectrum}

Brenner's $\kappa$ (kappa) estimate \citep{Brenner2010} (based on \cite{Robbins}) is $\hat{\pi}_q=(1{-}\kappa)/n$ when $k_q=0$, where $\kappa$ is the fraction of singleton profiles (observed only once in the database). Intuitively, a large $\kappa$ corresponds to a high profile diversity which justifies a low estimate for $\pi_q$. The estimator can be very small if the database consists mainly of singletons, which is the case for high-mutation-rate Y profiles. In particular, if the database consists only of singletons, then $\kappa = 1$ and $\hat{\pi}_q = 0$.

Cereda's estimators \citep{Cereda2017, Cereda2017b} are based on the numbers of both singletons and doubletons (profiles observed exactly twice). All of these approaches take no account of individuals with profile $q$ due to relatedness with $Q$, and the estimates can be strongly influenced by the way the database is sampled, as discussed above.

\subsubsection{Augmenting the database}
\label{sec:augment}

If we compute the database relative frequency after adding $q$, then both $k_q$ and $n$ are augmented by one to obtain $\hat{\pi}_q=(k_q{+}1)/(n{+}1)$. If we add two copies of $q$ to the database, corresponding to the profiles of both $Q$ and $X$ under $H_X$, we obtain \citep{WoE1st}
\begin{equation}
\hat{\pi}_q=\frac{k_q+2}{n+2}.\label{eq:n2}
\end{equation}
Use of \eqref{eq:n2} is conservative in the sense that we don't know if we have observed one or two individuals with $q$ (that is, we don't know if $H_X$ is true). Both the above estimators can alternatively be derived as Bayesian posterior probabilities given a uniform prior for $\pi_q$, the first using the original database while \eqref{eq:n2} uses the database augmented with one copy of $q$.

Other methods can be modified similarly by augmented the database with one or two copies of $q$.

\subsubsection{Upper confidence limit (UCL)}
\label{sec:ucl}

An alternative to \eqref{eq:n2} is an upper confidence limit (usually 95\%) for $\pi_q$ \citep{holland1999,budowle:ishi18}. This is the smallest binomial ``success'' probability $\pi$ such that the probability of observing up to $k_q$ successes in $n$ trials is $\le 0.05$. That is, the UCL is the largest $\pi$ such that
\begin{equation}
\sum_{x=0}^{k_q}{n\choose x} \pi^x(1{-}\pi)^{n-x}\le 0.05\label{eq:CP}
\end{equation}
which is sometimes called the Clopper-Pearson formula \citep{Clopper1934}. The UCL represents a standard scientific approach to controlling the risk of overstating the evidence: it provides an answer to the question of how big the unknown $\pi_q$ could reasonably be, given observations $k_q$ and $n$. The 95\% UCL is larger (more conservative) than \eqref{eq:n2}. For example, when $k_q=0$, from \eqref{eq:CP} and ln$(0.05)\approx-3$, we obtain that the UCL is just under $3/n$, whereas \eqref{eq:n2} is just under $2/n$.

\subsection{The Discrete Laplace method}
\label{sec:disclap}

The Discrete Laplace method \citep{AndersenDisclap2013, jossdisclapmix} models $\pi_q$ for all possible $q$, making it also useful for model-based clustering and mixture analyses \citep{AndersenDisclapCluster2014, AndersenDisclapMixture2015}. Intuitively, based on the database profiles, clusters are identified  corresponding to the descendants of a recent common ancestor, and probabilities are computed for the observed profiles to belong to each cluster. The ancestors are treated as unrelated and so their profiles are independent. Profile probabilities are computed by assuming that profiles descend independently from each ancestor according to a discrete Laplace (double geometric) distribution.

\subsection{Coancestry adjustment for population sub-structure}
 \label{sec:theta}

The population genetic parameter $F_{ST}$, often referred to as $\theta$ in forensic DNA profiling, has been widely used to correct match probabilities for autosomal DNA profiles \citep{BaldingNichols1994}. An analogous adjustment for lineage marker profiles \citep{WoE2nd,Buckleton2011} is
\begin{equation}
\mbox{LR}=\frac{1}{\theta + (1{-}\theta) \pi_q}\label{eq:theta}
\end{equation}
There are several interpretations of $\theta$, one is that it represents the average level of relatedness of individuals within a subpopulation, relative to a larger population. In forensic settings the larger population can be interpreted as the population from which the available profile database was drawn, while the subpopulation is not usually well defined but it is assumed to include $Q$ and some or all of the alternative possible sources of the evidence sample. The denominator of \eqref{eq:theta} can be loosely interpreted as follows. Under $H_X$, there is probability $\theta$ that $X$ is a relative of $Q$, sufficiently close that a match is very likely, while with probability $1{-}\theta$, $X$ comes from the broader population which includes more distant relatives of $Q$ such that the match probability is well estimated by the database relative frequency.

Equation \eqref{eq:theta} can be used in conjunction with the Discrete Laplace or other method to estimate $\pi_q$.  However, it is difficult to choose a $\theta$ value relevant to a particular case, which cannot be directly estimated. It will depend on the available case-specific background information and the population that the reference database was drawn from. In general, if the database is drawn from a broad, heterogeneous population that extends greatly beyond the population relevant to the case, then a larger $\theta$ would be needed than if the database came from a smaller more homogeneous region that includes most alternative donors of the DNA in a particular case.

Population-genetic estimates of $\theta$ have been reported for many human subpopulation-population pairs \citep{SteeleAHG2014,BuckFSI2016} and from simulation scenarios \citep{SteeleSJ2014}. Most available estimates of $\theta$ are for autosomal loci. For lineage markers, the smaller effective population size (there are 4 copies of each autosome for every Y chromosome) tends to increase the value of $\theta$, but the higher mutation rate of the Y tends to act in the opposite direction. Many population genetic studies do not estimate $\theta$ relative to a forensic database, but instead use a hypothetical ancestral reference population which leads to smaller estimates.  Further, few studies are available on the fine geographical scale relevant for many cases. 

In general, every alternative $X$ can have a different value of $\theta$ reflecting their level of ancestry shared with $Q$ relative to the database population. However, rather than try to choose a distribution of $\theta$ values appropriate for the alternative contributors in a case, usual practice is to take a single value from the upper tail of that distribution.  \citep{SteeleAHG2014} argued for $\theta=0.03$ as a conservative, default value for autosomal profiles, finding through simulations based on real data that it remains appropriate even if $X$ in fact comes from a different continent than the one where the database was sampled.

\subsection{Estimating the number of matches in the population}
\label{sec:numb-match}

If we knew that $K_q$ individuals 
have a profile matching $Q$, and these are well-mixed in a population of alternative sources of the DNA of size $N$, then the LR would equal the inverse of the match probability, which is
\begin{equation}
\mbox{LR} = \frac{N}{K_q}.
\end{equation}
We have noted above that this is not directly useful in practice, in contrast with the central role of the LR in the interpretation of autosomal profiles. This is because the choice of population and hence the value of $N$ is problematic for lineage marker profiles. As we consider larger suspect populations, $K_q$ increases at a slower rate than $N$, because the matching individuals are relatives of $Q$ and they are expected to form a smaller proportion of the population when it is more broadly defined.

An alternative to reporting the LR is to report estimates of $K_q$. A precise estimate is not usually possible, but probability distributions for $K_q$ can be obtained using simulation under different population genetic models \citep{AndersenPLOSGEN2017, AndersenPLOSGEN2018, AndersenYmix2019}, which can include various mutation rates and mechanisms, and demographic factors such as population growth and structure, as well as between-male variance in offspring number. The simulations can also condition on $k_q$ and $n$ from a database, provided the database sampling scheme is known, which is only feasible in practice if the database can be assumed to be a random sample from the population. Often conditioning on $k_q$ and $n$ doesn't greatly alter the distribution of $K_q$, since it merely confirms that the profile is rare as is expected from the mutation rate \citep{AndersenPLOSGEN2017}.

A further advantage of the simulation approach is that it can easily incorporate available information about the numbers of close relatives of $Q$, and about their profiles if available \citep{AndersenYmix2019}.

Using the malan (MAle Lineage ANalysis) software \citep{malan}, the distribution of $K_q$ was found in high-mutation-rate settings to be insensitive to the modelling assumptions \citep{AndersenPLOSGEN2017, AndersenYmix2019}. In the case of Yfiler Plus, the number of matching males is typically $<10$ and rarely more than a few tens. This approach has been extended to mitogenomes \citep{AndersenPLOSGEN2018}, but due to their lower mutation rate the distribution of $K_q$ was spread over a wider range, and was more sensitive to the population genetic model. While there remains merit in reporting to jurors an estimate of $K_q$, the arguments for this are less compelling than in high-mutation-rate settings, and conversely problems with the match probability and LR are lessened.

Given the estimates of $K_q$, a juror can assess how likely it is that one of these matching individuals was the source of the evidence profile, rather than $Q$. The population size $N$ may play some role in these considerations, but is relatively unimportant. Since $K_q$ is a count, it is likely to be more interpretable to a juror than an LR or a match probability \citep{Giger2000}: it can be presented using phrases such as "the number of individuals with profile $q$ is unlikely to be more than...".

\subsection{Methods not widely used}

\subsubsection{Frequency surveying}
\label{sec:frequency-surveying}

The frequency surveying method \citep{Roewer2000, Krawczak2001, Willuweit2011} is based on pairwise distances, measured in mutational steps, between the Y-profiles from individual $i$ and $j$. An exponential regression is then based on these distances and used to establish a Beta prior distribution in a Bayesian model with a binomial likelihood. One disadvantage of this approach is that the differences in mutation rates across Y-STR loci are ignored, and only counting the number of mutational steps discards information. This method is still available at \url{www.YHRD.org} \citep{YHRD15}, but is no longer recommended \cite{ISFG2020}.

\subsubsection{Population genetic modelling using the coalescent}
\label{sec:coalescent}

The first method for computing match probabilities using an explicit genetic model was based on a genealogical tree with $n{+}2$ leaves, representing the database augmented with one copy of $q$ plus a leaf node representing an unobserved profile \citep{Wilson2003}. Given a mutation model and a demographic model describing the population size, growth rates and structure, a Markov chain Monte Carlo algorithm updates the tree structure and branch lengths, as well as the profiles at the internal nodes of the tree and the unobserved leaf node. The distribution of profiles in the population is estimated from equilibrium frequencies at the unobserved leaf. The method was found to perform well in comparison to methods available at the time \citep{AndersenCoalescent2013}, but it is computationally demanding, particularly for large databases. This is because the whole profile space is explored, whereas only the probability of $q$ is needed for forensic identification.

\subsubsection{Graphical statistical models}

These Y-profile models \citep{Andersen2018, Andersen2020} are computationally fast and allow intermediate alleles, which the population-genetic models above cannot accommodate. However, they do not exploit much genetic information and alleles are merely assumed to be different categories.

\section{Recommendations from forensic authorities}
\label{sec:rec}


For Y-profiles, the Discrete Laplace method (Section \ref{sec:disclap}) is currently recommended in the Philippines \citep{RecomPhilip} and in Germany \citep{RecomGermany}, where it was first used in court in a case from 2015 \citep{Roewer2019}. In that case profiles were available for some male-line relatives of the suspected contributor, which can be taken into account in assessing the strength of the evidence by the malan simulation approach \citep{AndersenYmix2019} (Section~\ref{sec:numb-match}). The Polish Speaking Working Group of the International Society for Forensic Genetics (ISFG) \citep{RecomPoland} currently recommends using the $\kappa$ method (Section \ref{sec:spectrum}). The UK Forensic Regulator has recently commended use of \eqref{eq:n2}, the adjusted database frequency \citep{RecomUK}.

Below we briefly summarise and comment on recommendations from two other forensic authorities.

\subsection{Scientific Working Group on DNA Analysis Methods (USA, Canada)}

\subsubsection{Y-profiles}

The SWGDAM 2014 guidelines \citep{RecomSWGDAMY} note that ``the profile probability is not the same as the match probability'', but they do not distinguish between low- and high-mutation-rate profiles. They recommend that the profile probability is estimated by the unadjusted database relative frequency $k_q / n$, or with a UCL (Section~\ref{sec:ucl}), and that this value be used within \eqref{eq:theta} to adjust for population structure (Section~\ref{sec:theta}). 

The guidelines also discuss the importance of identifying the relevant population(s) and the difficulty in choosing an appropriate $\theta$ value, and they suggest default values for three kits. For PowerPlex Y23, the most discriminatory (highest profile mutation rate) of the three, they suggest $\theta=2\times10^{-5}$ for African Americans, Asians, Caucasians and Hispanics; and $\theta=3\times10^{-4}$ when Native Americans are considered.

\subsubsection{mtDNA}

The SWGDAM 2019 guidelines for mtDNA \citep{RecomSWGDAMmtDNA} are similar to those for Y-profiles \citep{RecomSWGDAMY}, except for the $\theta$ adjustment. They write: ``\textit{It is recognized that population substructure exists for mtDNA haplotypes. However, determination of an appropriate theta ($\theta$) value is complicated by the variety of primer sets, covering different portions of HV1 and/or HV2, which may be applied to forensic casework. SWGDAM has not yet reached consensus on the appropriate statistical approach to estimating $\theta$ for mtDNA comparisons}''.
 
\subsection{International Society for Forensic Genetics (ISFG)}

\subsubsection{Y-profiles}

The ISFG 2020 guidelines \citep{ISFG2020} state that ``Information on the degree of paternal relatedness in the suspect population as well as on the familial network is however needed to interpret Y-chromosomal results in the best possible way'' but did not delineate how to achieve this. The guidelines focus on estimating a profile relative frequency from a database, 
for which they recommend the Discrete Laplace method (Section~\ref{sec:disclap}), which we support in low-mutation-rate settings but the guidelines do not distinguish low- and high-mutation-rate settings.

The guidelines also recognise arguments for avoiding profile frequencies, referencing \citep{AndersenPLOSGEN2017}, but they write ``However, the appropriate wording for statements not relying on population databases need to be validated in the context of the national guidelines.''.

\subsubsection{mtDNA}

The ISFG 2014 guidelines on mtDNA \citep{ISFGmtDNA} focus on profile frequencies, but do not recommend any particular approach; instead they mainly mention frequency-based estimators such as \eqref{eq:n2} (Section~\ref{sec:augment}) and the UCL (Section~\ref{sec:ucl}).







\section{Some further issues}
\label{sec:fur}

\subsection{Combination with autosomal evidence}

In many respects, lineage marker profiles resemble a single locus of an autosomal profile, with a higher mutation rate and hence greater allelic diversity than for typical autosomal loci but only one allele rather than two. The problem of the strong role for relatedness that we have highlighted in this review also arises at a single autosomal locus, although due to diploidy there are four lineages connecting $Q$ and $X$ at an autosomal locus, one for each pairing of an allele from $Q$ with one from $X$.

It is possible to compute a combined LR for autosomal and lineage marker profiles that both match $Q$. This is rarely done in practice, perhaps because an autosomal profile match is typically so informative that the relatively small additional evidential strength of the lineage marker match is outweighed by the additional interpretation issues. 
One possible approach would be to multiply a conservative estimate of $K_q$ by the autosomal match probability, to obtain an estimate of the expected number of individuals matching $Q$ at both lineage marker and autosomal profiles. Alternatively, if the lineage marker profile mutation rate is low, it may be acceptable to multiply autosomal and lineage-marker LRs, each obtained using a suitable value of $\theta$.

\subsection{Locus order for duplicated Y loci}

The STR loci that form a Y-profile usually have known locations on the Y chromosome, except for some pairs of duplicated loci including DYS385a and DYS385b. Two males with the same pair of alleles DYS385a/b may or may not match because of the two possible orderings. When the two alleles from $Q$ are the same as those from the evidence profile, the evaluation problem can be overcome by ignoring the duplicated loci, which tends to underestimate evidential strength, or by assuming a match at both loci, which tends to overstate evidential strength, by at most a factor of two and arguably much less.

\subsection{Consistency}
\label{sec:consistency}

\citep{Cowell2020} proposes a consistency principle requiring that the strength of evidence for a Y profile cannot be less than that for any the sub-profiles obtained by omitting one or more loci. This reasonable requirement is not enforced by most of the methods discussed here, because a Y profile is treated as a single allele and sub-profiles are not considered. The problem can be important when there is a good sample size for a particular DNA profiling kit, but a reduced sample size for a more detailed profile that includes additional loci \citep{RecomSWGDAMY,RecomUK}. The method of Section \ref{sec:numb-match} based on the distribution of the number $K_q$ of matching individuals does respect the principle, because adding additional loci to a profile increases the mutation rate, and hence stochastically reduces $K_q$.

\subsection{Partial profiles}
 \label{sec:partial-profiles}
Often the DNA from a contributor of interest is at a very low level and/or degraded, so that a partial profile can arise if no allele is observed at some Y-STR loci, due to allele drop-out and/or masking by the alleles of a known contributor. The principles of interpretation are unchanged, using only the loci at which an allele is observed. The fewer the loci observed, the lower the profile mutation rate, which increases the number of matching individuals and decreases their average relatedness with $Q$. These quantities can be assessed using simulation including only the observed loci \cite{malan,mitolina}.

\subsection{Mixtures}
\label{sec:mixtures}

Many evidence profiles come from multiple contributors, with at least two of them unknown. If there is a large discrepancy in the amount of DNA from each unknown contributor, it may be possible to deconvolve the mixture (assign alleles to distinct, unknown contributors) manually, based on peak heights. Various peak height models are available. Mixture examples with peak height information are included in \citep{Taylor2018}.

Surprisingly, for high-mutation-rate Y-profiles, a two-male mixture profile in which $q$ is fully represented is almost as powerful as observing $q$ as a single-source evidence profile \cite{AndersenYmix2019}. Although many pairs of possible Y-profiles could give rise to the observed mixture, the great majority of these possible profiles do not exist in the worldwide human population, which is minuscule compared with the vast number of possible profiles. Since we know that $q$ does exist in the population, the pair of profiles that includes $q$ is much more likely than all the other profile pairs combined, unless one or more of the other profiles has been observed in a database \cite{AndersenYmix2019}.

For low mutation rate Y-profiles, the Discrete Laplace method (Section \ref{sec:disclap}) can be used to deconvolve mixtures using estimated population frequencies \citep{AndersenDisclapMixture2015}.

A "profile centred'' (HC) method \citep{Taylor2018} is based on \eqref{eq:mp-meioses} and focusses on the number of generations $G$ in the lineage linking $Q$ with alternative contributor $X$. Beyond some threshold on $G$, the method assumes that the match probability is low and uses a population frequency estimator similar to the Discrete Laplace method. The HC method assumes a constant-size, random mating population of size $N$ to assign weights for the mixture donors (using formulas for the probability that two random persons are $g$ generations apart). The HC method uses the $\kappa$ method (Section \ref{sec:spectrum}) for calibration, so that these methods are the same in the special case of good-quality single-source profiles.


\section{Discussion}
\label{sec:dis}

Our review of methods for assessing the forensic value of DNA evidence from lineage markers, namely Y-profiles and mitogenomes, has emphasised the key roles of the profile mutation rate and the (male-line or female-line) relatedness $G$, which are not adequately addressed by many methods. In particular, it is unsatisfactory to inform courts that, for example, a $Y$-profile match is likely for male-line relatives and then to proceed as if the alleged contributor $Q$ is unrelated to the alternative contributors.

Values of $G$ are typically unknown in actual populations except when $G$ is very small, but the distribution of $G$ can be investigated via simulation in population genetic models (Section \ref{sec:numb-match}). There are also some well documented actual human populations that can be studied in more detail. To date these are relatively isolated and not typical of many cosmopolitan urban populations. A national-scale project is underway in Denmark aimed at tracing lineages over a century for almost the whole population \citep{NNF}. At this time-depth, lineage paths up to about $G=8$ for two contemporary young adults can be traced, provided that there are no migrants in the lineage.

When the profile mutation rate is low (say below 0.05 per generation, which holds for the mitogenome and older Y-STR profiling kits), most of the individuals with profiles matching $Q$ are separated by at least several tens of generations. There are typically hundreds of matching individuals, and in some cases thousands of them \citep{AndersenPLOSGEN2018}. That is enough individuals and at sufficient genetic distance that many of them will differ from $Q$ in many characteristics, thus lessening the problems discussed above. In these cases, methods based on estimating population match probabilities such as the Discrete Laplace method recommended by the ISFG (Section \ref{sec:disclap}) may be acceptable, provided that the role of relatedness is also adequately explained possibly through a $\theta$ adjustment.  As discussed in Section \ref{sec:theta}, it is difficult to recommend specific values for $\theta$ as these can depend on case-specific details and the reference database.

For high-mutation-rate profiles (say above $0.1$ per generation) most males matching $Q$ are related to him within a few generations, and there seems no satisfactory alternative to summarising to a court the distribution of the number of close relatives of $Q$ expected to match, as well as their degree of relatedness (Section \ref{sec:numb-match}). The UCL, for example, will be conservative if the alternative sources of the DNA include few close relatives of $Q$, but without addressing that question directly we can't be sure. Methods that rely on a database may be affected by a non-conservative bias because the database is typically drawn from a broadly-defined population that has a lower average relatedness with $Q$ than the alternative sources of the DNA in a particular case.

To improve the simulation approaches to approximating the number of matches, we require good models and rate estimates for mutation. Some data are available but details such as the dependence of mutation rate on current profile state have been little studied, particularly for mitogenomes. Mutation models allowing for allele specific mutation rates have been investigated \cite{Jochens2011}, but the available data was insufficient to estimate the model parameters accurately.

\bibliography{references}

\end{document}